\documentclass[lettersize,journal]{IEEEtran}
\usepackage{amsmath,amsfonts}
\usepackage{algorithmic}
\usepackage{algorithm}
\usepackage{array}
\usepackage{textcomp}
\usepackage{stfloats}
\usepackage{url}
\usepackage{verbatim}
\usepackage{graphicx}
\usepackage{cite}
\usepackage{caption}
\usepackage{makecell}
\usepackage{multirow}
\usepackage{subfigure}
\usepackage[export]{adjustbox}
\usepackage{soul}
\usepackage{ulem}

\usepackage{threeparttable}

\begin{document}

\title{{Near-Field Channel Characterization for Mid-band ELAA Systems: Sounding, Parameter Estimation, and Modeling}}

\author{Wei Fan, Zhiqiang Yuan, Yejian Lyu, Jianhua Zhang, Gert Pedersen, Jonathan Borrill, and Fengchun Zhang

\thanks{Wei Fan is with Southeast University, China; Zhiqiang Yuan and Jianhua Zhang are with the Beijing University of Posts and Telecommunications, China; Yejian Lyu, Gert Pedersen, and Fengchun Zhang are with Aalborg University, Denmark; Jonathan Borrill is with Anritsu, Sweden.}}

\maketitle

\begin{abstract}

6G communication will greatly benefit from using extremely large-scale antenna arrays (ELAAs) and new mid-band spectrums (7-24 GHz). These techniques require a thorough exploration of the challenges and potentials of the associated near-field (NF) phenomena. It is crucial to develop accurate NF channel models that include spherical wave propagation and spatial non-stationarity (SnS). However, channel measurement campaigns for mid-band ELAA systems have rarely been reported in the state-of-the-art. To this end, this work develops a channel sounder dedicated to mid-band ELAA systems based on a distributed modular vector network analyzer incorporating radio-over-fiber (RoF), phase compensation, and virtual antenna array schemes. This novel channel-sounding testbed based on off-the-shelf VNA has the potential to enable large-scale experimentation due to its generic and easy-accessible nature. The main challenges and solutions for developing NF channel models for mid-band ELAA systems are discussed, including channel sounders, multipath parameter estimation algorithms, and channel modeling frameworks. Besides, the study reports a measurement campaign in an indoor scenario using a 720-element virtual uniform circular array ELAA operating at {16-20} GHz, highlighting the presence of spherical wavefronts and spatial non-stationary effects. The effectiveness of the proposed near-field channel parameter estimator and channel modeling framework is also demonstrated using the measurement data.

\end{abstract}


\section{Introduction}
\label{sec:intro}
Multi-antenna technology has found its wide applications in various wireless standards and commercial radio products, ranging from 4G long-term evolution (LTE), and 5G new radio (NR) to WiFi technologies. It is expected that future base stations beyond 5G will continue to evolve to accommodate extremely large antenna array (ELAA) configurations. The integration of ELAA systems has the potential to significantly improve the wireless system performance. Thanks to the smaller wavelength at higher frequency bands and a larger physical antenna aperture, future base stations can accommodate many more radiating antennas, which can provide high array gain and extremely narrow beams to combat propagation loss and improve energy and spectral efficiency\cite{bjornson2016massive}.  

Moreover, the upcoming 6G era necessitates new considerations in frequency spectrum allocation to facilitate various 6G use cases, spanning from omnipresent connectivity and massive machine-type communications to extreme data-rate transmission scenarios. The spectrum below sub-6 GHz can typically address use cases for wide-area coverage and high mobility, while the mmWave range spectrum can provide high capacity in short-range populated environments. Different spectrum ranges, including the existing spectrum and new spectrum in the essential mid-band ({7-24} GHz) and the supplementary sub-THz ({92-300} GHz) will become necessary in 6G. The mid-band range, also unofficially termed as frequency range (FR) 3 in the industry, offers sufficient unexploited frequencies for wide bandwidth utilization compared to FR1 (up to 7.125 GHz) and wide coverage compared to FR2 (starting from 24 GHz). Therefore it has attracted huge interest from industry \cite{qual_tr} and standardization (e.g. 3GPP Release 19) in recent years. Radio technologies, employing ELAA with analog, hybrid, and digital beamforming structures covering the FR1, FR2, and FR3 bands, are increasingly emerging for future 6G communication.  

\begin{figure*}
    \centering
    \includegraphics[width=1.98\columnwidth]{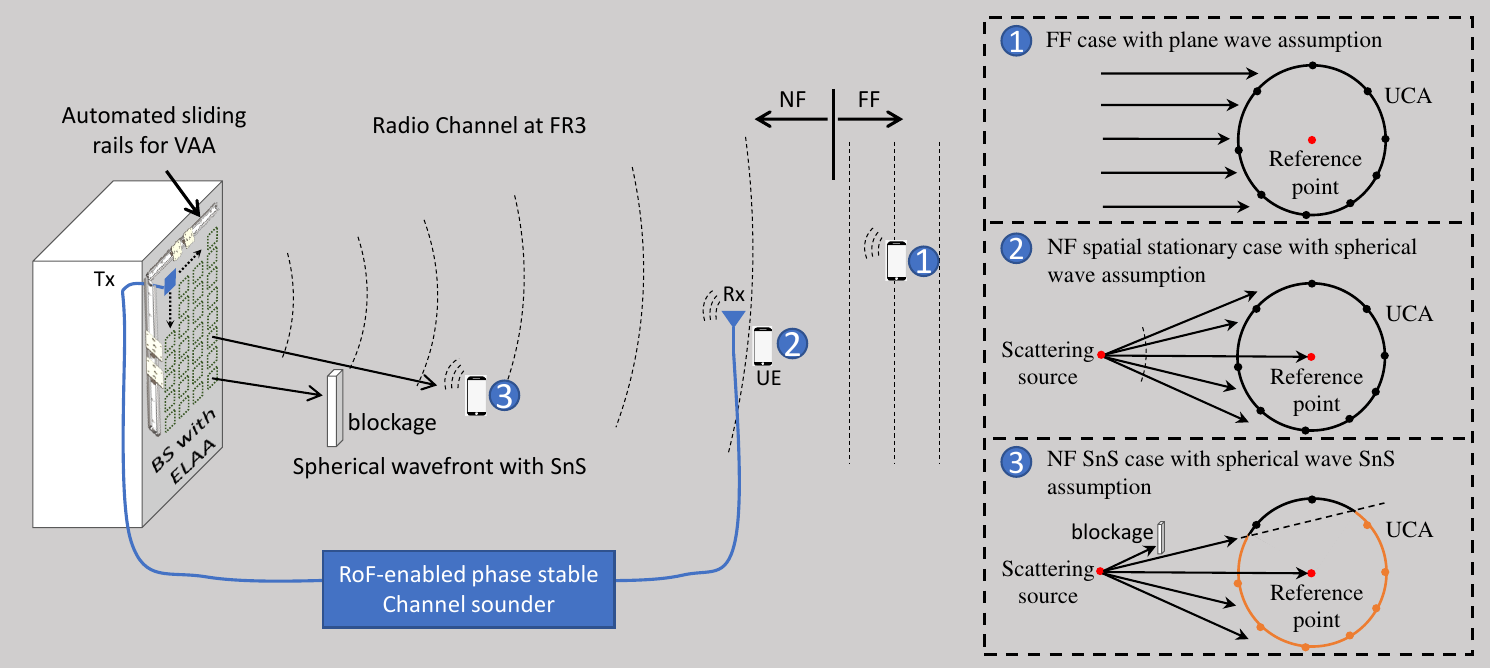}
    \caption {An illustration of VAA-based channel characterization for FR3 ELAA systems accounting for the FF plane-wave, NF spatial stationary, and NF SnS effects.}
    \label{fig:vaa_concept}
\end{figure*}

The radiation field of an antenna system can be divided into two main regions, the far-field (FF) and the near-field (NF), as depicted in Fig. \ref{fig:vaa_concept}. The boundary between these two regions is determined by the Fraunhofer distance, which is proportional to the array aperture size and operating frequency. In the FF region, plane-wave propagation and spatial stationarity are generally assumed. However, with ELAA systems operating at the higher mid-band range, the NF phenomena arise and have been widely observed, e.g. \cite{Fredric_sounder}, as illustrated in Fig. \ref{fig:vaa_concept}. Specifically, multipath impinges onto arrays with spherical wavefronts and spatial non-stationarity (SnS), i.e., array elements in the ELAA experience different channel characteristics.
Notably the spherical wave propagation and SnS are inherent properties of NF ELAA systems \cite{yuan2022spatial}.
The MIMO communication performance can benefit from NF properties that are inherently present in ELAA systems \cite{bjornson2016massive,heathsnsbeamforming}. For example, within the NF region, spherical waves can lead to a higher rank for the MIMO channel matrix, consequently achieving a higher degree of freedom compared to the respective FF scenario. The ELAA-type BSs can also exploit NF regions to help separate closely located spatial users, thus improving multi-user multiplexing performance\cite{heathsnsbeamforming}.  

To fully exploit the anticipated benefits of ELAA systems for near-field communications, it is essential to build accurate and realistic NF channel models for ELAA systems at the new deployment frequency bands, accounting for spherical waves and SnS effects within the existing channel modeling framework. However, due to the lack of practical ELAA channel sounding systems at the mid-band range, few works have been reported at the state-of-the-art, to our best knowledge. For this reason, a channel sounder dedicated to ultra-wideband ELAA systems is developed in this work based on a newly available long-range distributed modular 2-port vector network analyzer (VNA). The ELAA system is implemented based on the virtual antenna array (VAA) concept, which is a popular channel sounding scheme due to its low cost, flexibility, and simplicity.  In this work, we first introduce the key challenges and solutions for channel characterization of ELAA systems, focusing on channel sounding systems, parameter estimation algorithms, and channel modeling frameworks. After that, a measurement campaign in an indoor scenario, employing a 720-element (virtual) uniform circular array (UCA) ELAA system at {16–20} GHz, was reported, highlighting the presence of NF effects. A novel maximum likelihood estimation (MLE) algorithm tailored to capture the NF phenomena (both spherical wavefront and SnS effects) is reported to extract the path parameters from the recorded channel data. 

This work is structured as follows. In Section II, we review the recent works and challenges in ELAA channel measurements and characterization. In Section III, we present experimental characterization of the ELAA channel at the mid-band based on the developed channel sounder, estimator, and modeling framework. Finally, Section IV concludes the paper. 

\section{Radio channel Characterization for ELAA systems: State-of-the-art}
\label{sec:review}

\subsection{Radio Channel Sounder}
A high-fidelity channel sounder is essential for recording reliable channel data to build realistic channel models. The radio performance of the measurement systems should be more demanding than the transceivers in the communication systems for flexible performance evaluation \cite{jesper2018raa}. For ELAA systems, the key challenge in channel sounding is to construct an experimental system that offers the extreme capability to record channel spatial profiles, considering NF effects. Depending on schemes for measuring channel spatial profiles, channel sounders can be categorized as follows: 
\begin{itemize}
    \item Real antenna array (RAA): utilizes one radio-frequency (RF) chain for each radiating antenna and records real-time channel responses for all ELAA antenna elements simultaneously \cite{jesper2018raa};
    \item Switched antenna array (SAA): employs multiple radiating antennas connected to only one RF chain via RF switches, and records channel responses for all ELAA antenna elements sequentially \cite{xuesong2023saa};
    \item Phased array: achieves electrical beam-steering via allocating FF beam-steering weights to antenna array elements \cite{derek2019pa};
    \item Directional scanning scheme (DSS): records channel spatial profiles by mechanically rotating a highly directional antenna \cite{molisch2023dss};
    \item VAA: mechanically moves a single antenna to different predefined spatial locations to mimic the ELAA systems, and records the channel responses for each virtual ELAA element sequentially \cite{medbo2012directional, Fredric_sounder, xuesong_estima, RSschmieder2019directional, Tongj_yin2017scatterer, gigantic_VAA}.
\end{itemize}
However, the requirements for measuring spatial channels for the ELAA systems incorporating NF effects have posed significant challenges for these schemes. A non-exhaustive state-of-the-art channel sounders with large-scale antenna configurations in recent years are listed in Table \ref{table:current_meas}. 

For RAA and SAA-based sounders, the cost and calibration complexity will scale up significantly with the increasing number of radiating antenna elements in the ELAA system. The beam-steering type schemes (i.e. the phased array and DSS schemes) might be unsuitable for NF channels since the channel responses for ELAA elements cannot be directly obtained. The VAA scheme has gained popularity in recent years due to its simplicity, low cost, and generality. It is capable of forming virtual ELAA systems with arbitrary configurations and covering various frequency bands in principle. Additionally, VAA eliminates antenna mutual coupling effects, which is beneficial for performing beamforming to extract propagation channel parameters. However, VAA-based measurements, though typically automated with the help of a control computer for mechanical systems, require mechanical movement to record channel response for each virtual ELAA element. As a result, it demands a significant amount of time, making it unsuitable for channel measurements in dynamic deployment scenarios.

  \begin{table*}
        \caption{State of the art for ELAA channel sounders.}
        \label{table:current_meas}
        \setlength{\tabcolsep}{5pt}
        \centering
        \renewcommand\arraystretch{1.5}
        \begin{tabular}{| c | c | c |c|c|}
        \hline
        \makecell[tc]{FR} & 
        \makecell[tc]{Research} & 
        \makecell[tc]{Sounder type} &
        \makecell[tc]{Frequency} &
        \makecell[tc]{Objectives and descriptions}\\
        \hline
        \multirow{2}{*}{FR1}&
        \makecell[tc]{\cite{Fredric_sounder}}&
        \makecell[tc]{128-ele. VAA-type ULA} &
        \makecell[tc]{2.6 GHz, 50 MHz BW}&
        \makecell[tl]{Preliminary observations of the spherical waves and SnS}\\
        \cline{2-5}
        {}&
        \makecell[tc]{\cite{jesper2018raa}}&
        \makecell[tc]{64-ele. S\&RAA-type UPA} &
        \makecell[tc]{5.8 GHz, 100 MHz BW}&
        \makecell[tl]{Massive MIMO channel measurements and performance analysis}\\
        \hline
        \multirow{4}{*}{FR2}&
        \makecell[tc]{\cite{xuesong2023saa}}&
        \makecell[tc]{$128$/$256$-ele. SAA-type UPA} &
        \makecell[tc]{27.5-29.5 GHz}&
        \makecell[tl]{Design and implementation of an SAA-based channel sounder}\\
        \cline{2-5}
        {}&
        \makecell[tc]{\cite{derek2019pa}}&
        \makecell[tc]{$8\times 32$ phased UPA array} &
        \makecell[tc]{62.5 GHz}&
        \makecell[tl]{Design of a phased array-based sounder and its validation}\\
        \cline{2-5}
        {}&\makecell[tc]{\cite{xuesong_estima}}&
        \makecell[tc]{360-ele. VAA-type UCA} &
        \makecell[tc]{27-29 GHz}&
        \makecell[tl]{Validation of an MLE algorithm for channel parameter estimation}\\
        \cline{2-5}
        {}&
        \makecell[tc]{\cite{RSschmieder2019directional}}&
        \makecell[tc]{1000-ele. VAA-type UCA}&
        \makecell[tc]{28 GHz, 2 GHz BW}&
        \makecell[tl]{Analyzing channel multipath characteristics for standardization}\\
        \hline
        \multirow{2}{*}{FR3}&
        \makecell[tc]{\cite{Tongj_yin2017scatterer}}&
        \makecell[tc]{$11\times11$ VAA-type UPA}&
        \makecell[tc]{9.25-9.75 GHz}&
        \makecell[tl]{Localization of scatterers in NF spherical wave propagation}\\
        \cline{2-5}
        {}&
        \makecell[tc]{This work}&
        \makecell[tc]{720-ele. VAA-type UCA}&
        \makecell[tc]{16-20 GHz}&
        \makecell[tl]{FR3 ELAA channel sounding, parameter estimation, and modeling}\\
        \hline
        \multirow{1}{*}{W-band}&
        \makecell[tc]{\cite{gigantic_VAA}}&
        \makecell[tc]{2400-ele. VAA-type UCA}&
        \makecell[tc]{95-105 GHz}&
        \makecell[tl]{Review on Sub-Terahertz channel characterization and an example}\\
        \hline
        \end{tabular}
        \begin{tablenotes}\footnotesize
            \item[+] + Abbreviation: UPA/ULA/UCA: Uniform planar/linear/circular array, ele.: element, S\&RAA: SAA-RAA, BW: bandwidth.
        \end{tablenotes}
        \label{table:summary}
        \end{table*}

Most VAA schemes reported in the state-of-the-art are implemented with the VNA. However, in conventional VNA systems, the signal generation and analysis units are co-located, making an extension cable essential to remote antennas for channel-sounding purposes in practical deployment scenarios. Coaxial cables are lossy, especially in the high-frequency bands, resulting in a short measurement range. Radio-over-fiber (RoF) scheme is an effective solution for achieving long-range measurements since it offers negligible signal loss over optical cables. Phase in the optical cable is susceptible to thermal change and mechanical stress, making accurate phase measurement challenging. A phase compensation scheme has been proposed to address the random phase issue, which basically employs a feedback link to record and correct the random phase in the forward link \cite{yuan2022spatial}. With these recent developments, we can employ the VAA scheme for channel sounding of ELAA systems, with frequency range up to sub-THz bands, and support for long measurement range. However, dedicated electrical-to-optical (E/O) and optical-to-electrical (O/E) units, bi-directional signal transmission, and post-processing calibration are required. Moreover, channel sounders and associated measurement campaigns for ELAA systems at the FR3 bands have been rarely reported. 

\subsection{Channel Multipath Parameter Estimation}

Multipath parameter estimation aims to extract multipath parameters from the measured channel responses by exploring the parameter spaces and finding the best match between the signal model and the observed channels. For ELAA systems, the use of gigantic antenna arrays can enhance channel spatial parameter estimation accuracy by leveraging the phase coherency of signals across multiple antenna elements \cite{bjornson2016massive}. However, the ELAA also introduces extra challenges due to the emerging NF effects. The spherical-wave propagation and SnS effects pose challenges to existing multipath estimation algorithms in terms of accuracy and complexity in the state-of-the-art\cite{xuesong_estima}.

The beamforming algorithms extract the path's spatial parameters by boosting the received signals at target directions via allocating a complex FF weighting vector. These methods are robust and computationally efficient due to their simplicity. However, beamformers typically assume a plane-wave model and spatially stationary propagation paths, which do not account for spherical waves and SnS present in NF ELAA channels. The sub-space algorithms decompose the channel covariance matrix into signal and noise sub-spaces, enabling joint angle and delay parameter estimation by exploiting the orthogonality between these sub-spaces. However, these techniques are often limited to narrowband and plane-wave propagation assumptions, which becomes problematic for wideband ELAA systems exhibiting NF properties. The MLE-based methods iteratively search for channel parameters that maximize the likelihood function between the measured and modeled channels. These methods are known for their high resolution, as well as high computational complexity due to joint parameter estimation and iterative searching processes. Representative methods, such as space-alternating generalized EM (SAGE) and Richter's maximum-likelihood estimation (RiMAX), aim to reduce the complexity by decomposing the multi-dimensional parameter search problems into several one-dimensional searches based on narrowband and FF assumptions. However, these assumptions may not hold in ELAA channels with NF properties. Some studies have developed MLE-based methods for NF cases by introducing the spherical-wave model but at the cost of significantly increased computational complexity.

In summary, the NF spherical-wave propagation effect causes accuracy degradation and complexity increase for current parameter estimation algorithms, and in addition, the SnS effect has been largely overlooked in the literature. Therefore, it is essential to consider these practical NF properties to accurately parameterize the ELAA channel. 

\subsection{Channel Modeling}

Channel modeling is the development of a framework that mathematically represents the propagation channel and describes how the transmitted signal is mapped to the received signal. Channel models should be accurate, and as simple as possible to enable low-cost implementation. For ELAA systems, the main challenge in channel modeling is to incorporate emerging NF spherical-wave propagation and SnS properties into existing modeling frameworks. Meanwhile, due consideration should be given to the additional complexity introduced by these new features.

Two popular channel modeling frameworks have been reported in the literature, i.e. statistical and deterministic models\cite{molisch2023dss}. For statistical channel modeling, the NF spherical-wave propagation can be characterized by a spherical-wave model with the introduction of scatterer distances. The SnS property, which manifests as path visibility/invisibility on elements and power variation on visible elements, can be captured through a combination of the visibility region (VR) concept (for path visibility/invisibility) and uniform theory of diffraction (UTD) theory (for power variation). In \cite{yuan2022spatial}, a statistical modeling framework that incorporates the spherical-wave model, VR concept, and UTD theory to characterize the NF SnS channel is proposed. The effectiveness of the framework has been experimentally validated. For deterministic channel modeling, such as ray-tracing (RT), both the NF spherical waves and SnS can be accurately characterized through a brute-force approach, i.e., simulating RT channels for each element pair between the transmitter (Tx) and receiver (Rx) arrays. However, the brute-force strategy suffers from high computational complexity due to the extensive RT simulations. A potential solution is to employ a coarse-to-refine simulation structure. Initially, RT is performed for sparsely located elements on both the Tx and Rx arrays, and then the RT-generated channels are used to interpolate the channels of the remaining element pairs based on spherical wave propagation and SnS capturing. In principle, this strategy can significantly reduce the complexity of RT simulations for ELAA channel modeling while maintaining accuracy. 

A key bottleneck in developing channel models for ELAA systems at the FR3 band is the lack of experimental data. Conducting extensive measurement campaigns in deployment scenarios is essential to develop realistic parameter distributions for the statistical channel models and to validate and calibrate the RT simulations. 

\section{VAA-based  Channel Characterization for ELAA systems at FR3} 
\label{sec:exploiting}
This section details the experimental characterization of the NF channels for the ELAA system at the FR3 using the VAA scheme. We developed a VAA-based channel sounder based on the distributed modular 2-port VNA to record the NF channels for the ELAA systems at the FR3 frequency band. Then, we present a newly developed channel parameter estimator that can characterize the NF effects. Finally, we reconstructed the channels for the ELAA system based on the estimated channel parameters and compared them with the measured data.

\subsection{Channel Sounding System for the ELAA system at FR3}\label{subsec:C_sound}

   \begin{figure}[!t]
        \centering
        \includegraphics[width=\columnwidth]{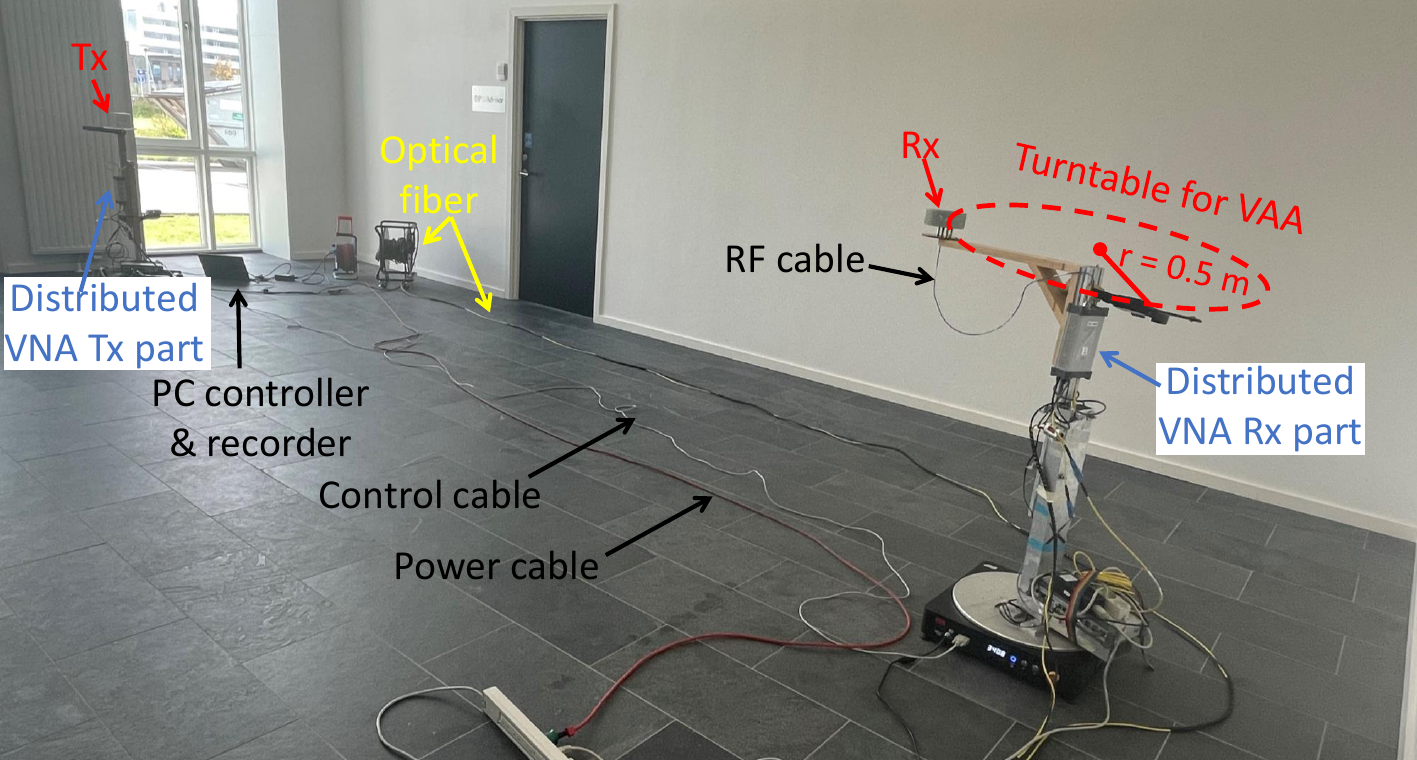}
        \caption{ Channel sounding system using the Anristu distributed Modular 2-port VNA.}
        \label{fig:VNA_sounder}
        \end{figure}

An off-the-shelf Anristu distributed Modular 2-port VNA (Anritsu ShockLine TM ME7868A series), which offers long range (enabled by the RoF solution) and phase-stable measurement capability (offered by the phase compensation scheme), was employed in the measurements. Unlike conventional VNA systems, the signal generator and analyzer units of this distributed VNA are distributed and connected by two optical fibers for signal transmission and synchronization, as illustrated in Fig. \ref{fig:VNA_sounder}. A phase compensation scheme is also implemented to ensure accurate and coherent phase measurement, which is required for achieving the VAA scheme. 

The channel measurement was carried out in an indoor line of sight (LOS) scenario, as shown in Fig. \ref{fig:VNA_sounder}. Two vertically polarized omnidirectional biconical antennas connected to the two distributed units of the VNA were employed as the Tx and Rx antennas, respectively. Both antennas were placed at a height of 1.25 m. On the Rx side, the VAA was formed by rotating the Rx antenna using a turntable, with a $50$ cm offset set intentionally to the rotation center. A rotation step of $0.5^{\circ}$ was set to cover the $360^{\circ}$, resulting in a virtual UCA with 720 elements. For each virtual UCA element, the channel frequency response (CFR) was recorded, covering {2-30} GHz with $14001$ frequency points. Given the focus on the FR3 band, the frequency range  {16-20} GHz with $2001$ frequency points was selected for study in this work. The FF distance of the UCA is 133 m at 20 GHz, which is significantly larger than the LOS distance of 6.7 m. The intermediate frequency bandwidth was set to $500$ Hz to ensure an excellent dynamic range. 

\begin{figure}
        \centering
        \includegraphics[width=\columnwidth]{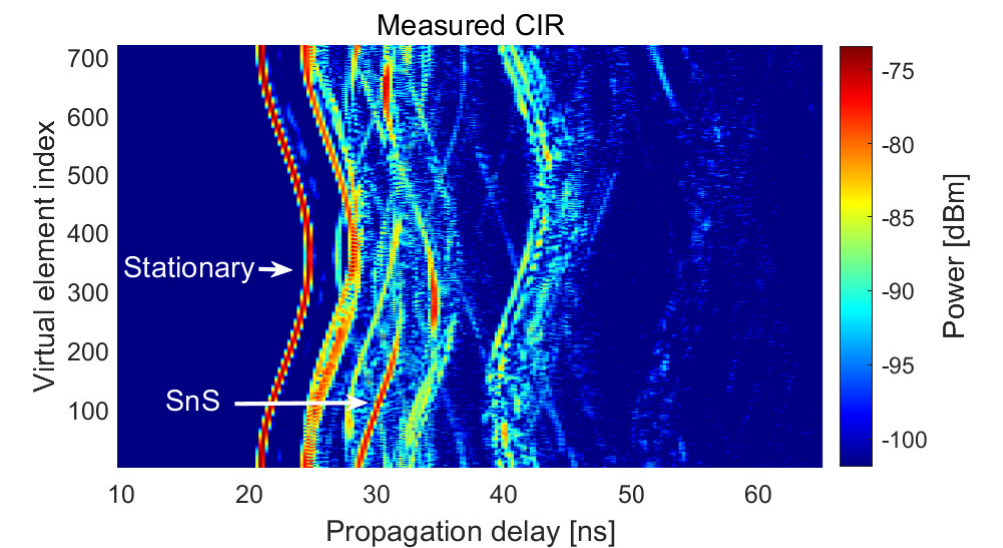}
        \caption{Measured CIRs across ELAA elements.}
        \label{fig:FR3_results}
        \end{figure}

After performing the inverse Fourier transform to the measured CFR for each UCA element, the channel impulse response (CIR) for the 720 virtual UCA elements can be obtained, as shown in Fig. \ref{fig:FR3_results}. The 's'-shaped curves are observable, denoting multipaths with different delays distributed on elements due to the large array aperture and spherical wave propagation. Furthermore, several complete trajectories (i.e. spatial stationary paths) and incomplete trajectories (i.e. SnS paths) can be identified. These observations indicate the presence of NF effects, i.e., both the spherical wavefront and the SnS effects, in the measured channels. Readers can refer to \cite{Fredric_sounder,haneda2006parametric} for theoretical analysis and experimental validation of the present NF phenomena.

\subsection{Channel Parameter Estimator for NF SnS channels}
\label{subsec:C_para}

There is a strong need to develop a low-complexity and generic channel estimator that can accurately estimate multipath parameters for both FF and NF channels, considering the presence of both spatial stationary and non-stationary paths. Several strategies can potentially address these requirements. 1) To capture the NF spherical-wave propagation, the spherical-wave signal model can be incorporated into traditional MLE algorithms by considering the distance between scatterers and transceivers. An MLE-based method incorporating spherical wave models has been developed for NF channel parameter estimation \cite{Tongj_yin2017scatterer}, albeit at the cost of increased computation complexity. 2) The SnS effect can be effectively captured using the VR concept to describe path visibility/invisibility and UTD for modeling power variation \cite{yuan2022spatial}. However, introducing these schemes will significantly increase complexity due to the additional parameter estimation associated with the spherical-wave model, VR concept, and UTD theory. 3) To mitigate complexity while maintaining its accuracy and effectiveness, a coarse-to-refine estimation framework can be employed. Low-complexity yet robust algorithms can be implemented initially to constrain the parameters within a coarse region. Subsequently, a refinement search can be conducted to optimize the estimator for achieving high resolution. This structure significantly reduces complexity while maintaining the parameter estimation accuracy. 

In this work, we propose a novel algorithm for multipath parameter estimation that incorporates the aforementioned strategies. We implement the algorithm based on our measured ELAA FR3 channel data as shown in  Fig. \ref{fig:FR3_results}. The algorithm follows the following steps. First, the spherical-wave model is introduced to capture the spherical-wave propagation, using both angles and spherical distance to characterize the spatial profile of a path. A non-stationary vector is defined for each path to represent its SnS behavior on the array elements. The vector entries are set within $\left[ 0,1 \right] $ where 0 constrains the path VR and $\left( 0,1 \right] $ denotes the power variation in the VR. Then, a robust beamformer \cite{xuesong_estima}, capable of achieving an approximately invariant beam pattern for both FF and NF scenarios, is adopted to coarsely estimate the path spatial and power parameters. Finally, an MLE-based procedure is exploited to search all parameters within the coarse range that maximize the likelihood function between the measured channel and the parameterized channel. {The proposed algorithm is advantageous from two aspects. In terms of accuracy, it is the first time that the channel SnS is adequately characterized in channel parameter estimation. Besides, in terms of complexity, we exploit a robust beamformer for initialization to narrow the path search regions, and decompose the high-dimensional joint parameter search into several low-dimensional searches based on the parameter orthogonality. These processes facilitate the cost-effective implementation of the MLE framework.}

\begin{figure}
          \centering
         \includegraphics[width=\columnwidth]{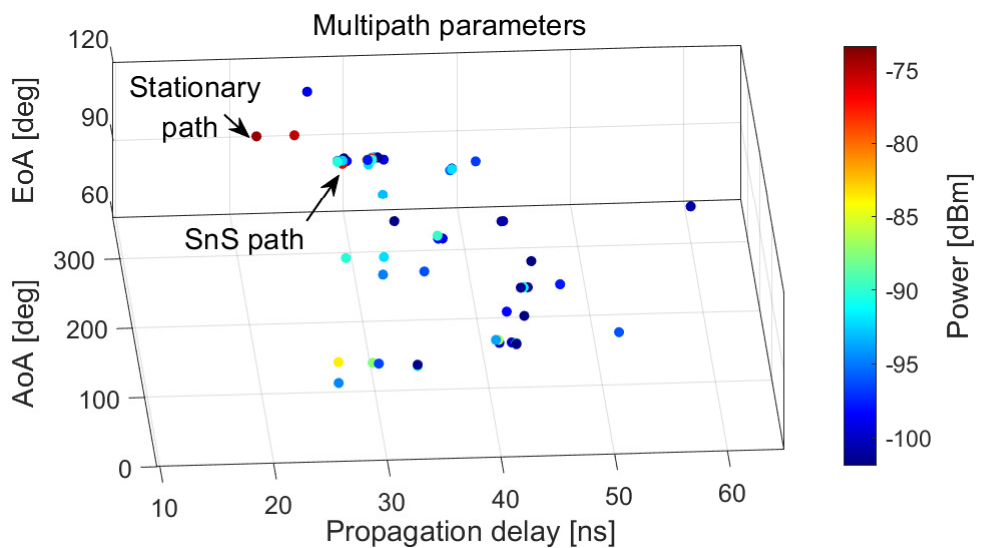}   \\
           \caption{MLE-based parameter estimation on the measured CIRs.}
        \label{fig: MLE results}
        \end{figure}

By applying the proposed algorithm to the measured ELAA FR3 channel data, in total 67 paths are extracted within a power dynamic range of 30 dB. The estimated path parameters are depicted in Fig. \ref{fig: MLE results}. Each data point in the figure represents an estimated path, with its position indicating the azimuth and elevation angles of arrival (AoA and EoA), along with the delay. The color of each point represents the corresponding path power. Those path parameters accurately reflect the location and shape of the curves in the measurement according to the channel representation \cite{yuan2022spatial}. Notably, the other two path parameters, i.e. the spherical distance and SnS coefficient vector that are estimated by the algorithm, are not demonstrated in the results in Fig. \ref{fig: MLE results} due to the page limit. 

The result in Fig. \ref{fig: MLE results} demonstrates that the algorithm is capable of distinguishing closely located paths with similar delays and angles, showcasing the algorithm's capability of high-resolution estimation. The estimated paths align well with the measured CIRs across the array elements. Specifically, each data point in Fig. \ref{fig: MLE results} corresponds to an 's'-shaped curve in Fig. \ref{fig:FR3_results}, and the estimated parameter values reflect the location (i.e., delays) and shape of the curves (angles and distances). Furthermore, the two paths marked in Fig. \ref{fig:FR3_results}, representing stationary and SnS trajectories, are successfully identified in the estimation result. Particularly, the SnS path corresponds to a path that undergoes an incomplete reflection by an elevator surface, highlighting the fact that certain objects with limited sizes may not act as complete scatterers for a large array in the NF region. The obtained path parameters will be utilized in the subsequent channel response reconstruction process to generate channels for various scenarios.

\subsection{Channel Modeling}
\label{subsec:C_mod}
\begin{figure}\centering
\includegraphics[width=\linewidth]{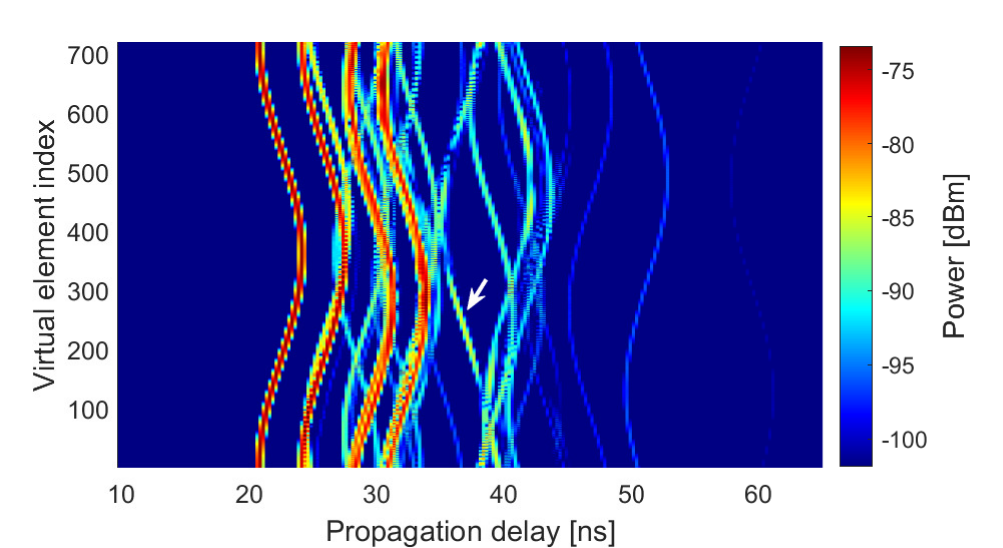} \\
\small (a) the FF case
\includegraphics[width=\linewidth]{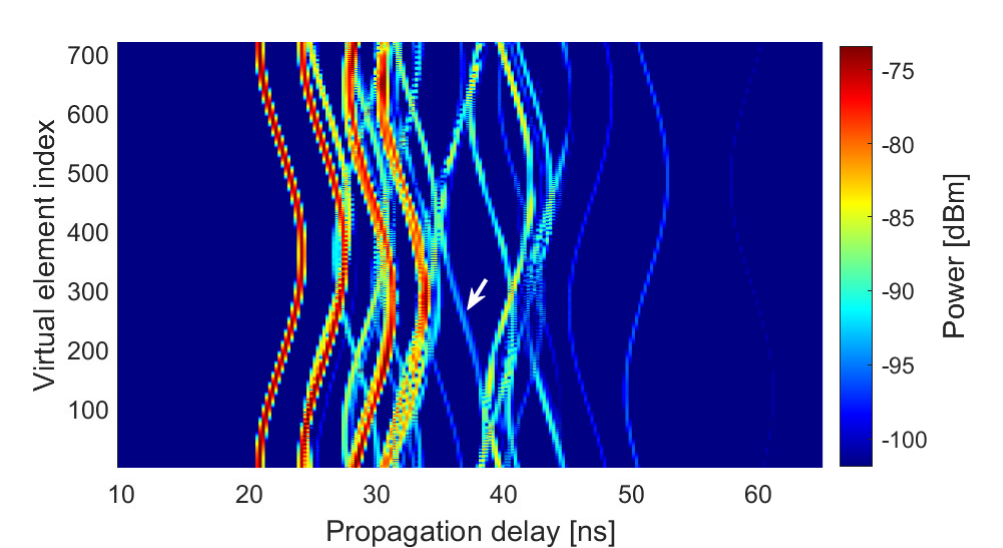} \\
\small (b) the NF stationary case
\includegraphics[width=\linewidth]{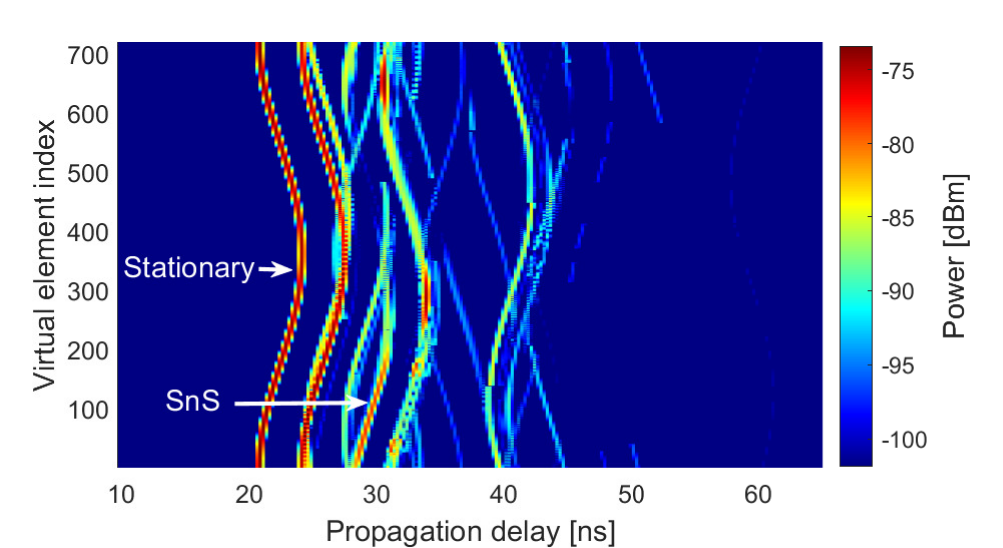}\\
\small (c) the NF SnS case
\caption{Generated CIRs in channel modeling with the FF, NF stationary, and NF SnS assumptions.}\label{fig:CIRs}
\end{figure}

We have implemented the statistical modeling approach for ELAA FR3 channels with a specific focus on capturing the NF spherical-wave propagation and SnS properties. Using the estimated channel parameters discussed in Section \ref{subsec:C_para}, i.e. AoA, EoA, propagation delay, power, and spherical distance and SnS coefficient vector for each path, we can reconstruct CIRs for the ELAA systems under different assumptions, i.e., FF, NF stationary, and NF SnS cases. 
Specifically, the NF SnS channel can be reconstructed as ${ \mathbf{H}^{\mathrm{sns}}\left( f \right) =\mathbf{S}\odot \mathbf{A}\left( f \right) \cdot \mathbf{H}\left( f \right) }$, with ${\mathbf{H}\left( f \right) }$, ${\mathbf{A}\left( f \right) }$, and ${ \mathbf{S}}$ denoting paths’ CFR at the reference point, the array manifold, and the SnS matrix, respectively\cite{yuan2022spatial}. The NF stationary channel can be generated by setting ${ \mathbf{S}}$ as the all-one matrix, and the FF channel can be reconstructed by further setting ${\mathbf{A}\left( f \right) }$ as the FF plane-wavefront manifold. Fig. \ref{fig:CIRs} depicts the reconstructed channels.

In the FF case, as shown in Fig. \ref{fig:CIRs}(a), paths are generated with complete trajectories on all array elements under the plane wave propagation and spatial stationary assumption. This assumption is commonly made in state-of-the-art algorithms. We can observe many complete 's'-shaped curves, each denoting a path with a plane wavefront. 

Under the NF stationary assumption, as illustrated in Fig. \ref{fig:CIRs}(b), paths are reconstructed with complete trajectories on a subset of array elements. The spherical-wave propagation in the NF region leads to phase responses of array elements different from those in the FF case. The marked example in Fig. \ref{fig:CIRs}(b) highlights the difference caused by complex-valued multipath decomposition with varying phase responses. The generated CIR accurately represents the NF stationary channel and can be used to evaluate the performance degradation of conventional multi-antenna technologies proposed under the FF assumption, as reported in \cite{haneda2006parametric}. Additionally, it serves as a guide for designing new NF communication technologies.

Fig. \ref{fig:CIRs}(c) shows the generated CIRs considering both the NF spherical waves and SnS. The modeled CIRs exhibit a high similarity to the measured CIRs in Fig. \ref{fig:FR3_results}, particularly for the dominant paths. The modeled CIRs accurately capture both stationary and SnS trajectories of paths. This demonstrates the accurate realization of the ELAA FR3 channel with the NF properties, providing a realistic platform for designing, evaluating, and testing FR3 ELAA technologies for NF communication with SnS.

\section{Conclusions}
NF phenomena introduced by the ELAA systems have brought new opportunities and challenges for next-generation wireless communications. Wireless communication channel testbeds are essential for the design and performance evaluation of radio systems with ELAA configuration at the promising mid-band frequency range. In this work, we briefly summarize the key challenges and state-of-the-art works in building such a channel testbed channel for the ELAA systems, with a focus on channel sounding, parameter estimation, and channel modeling. After that, we report a channel sounder based on the distributed module VNA, which is suitable for the ELAA antenna systems at the FR3 frequency band via employing the VAA scheme. A measurement campaign employing the developed channel sounder is carried out and realistic NF phenomena can be observed in the measured channel. A channel parameter estimator, which incorporates the NF spherical wave (by introducing the scatterer location) and the SnS effect (by incorporating the VR concept and UTD) is developed with complexity-reduction schemes. Based on the estimated multipath parameters, the channel responses for the ELAA are reconstructed under the FF, NF stationary, and NF SnS assumptions. The restructured CIRs incorporating the NF effects exhibit a high similarity to the measured CIRs of the ELAA systems, particularly for the dominant propagation paths. {The developed channel sounder and associated post-processes serve real-world experimental needs for the performance evaluation of ELAA mid-band technologies in typical and operational deployment scenarios.}

\section*{Acknowledgments}
This work was supported in part by the Anristu-funded project "Novel Channel Sounding Techniques for 6G", in part by the European Partnership on Metrology Project MEWS under Grant 21NRM03, and in part by the European 6G-SHINE project.

\normalem
\bibliographystyle{IEEEtranchange}
\addcontentsline{toc}{section}{\refname}\bibliography{Li_library}        


\bibliographystyle{IEEEtran}
\begin{IEEEbiographynophoto}{Wei Fan}  (weifan@seu.edu.cn) is currently a professor with Southeast University, China and a docent with Centre for Wireless Communications, University of Oulu. His current research interests include over-the-air (OTA) testing of multiple antenna systems, radio channel sounding, parameter estimation, modeling, and emulation.
\end{IEEEbiographynophoto}
\vfill

\begin{IEEEbiographynophoto}{Zhiqiang Yuan} (yuanzhiqiang@bupt.edu.cn) has been pursuing the Ph.D. degree in Beijing University of Posts and Telecommunications (BUPT) since 2020 and has been visiting Aalborg University since 2021. He currently studies channel sounding and characterization.
\end{IEEEbiographynophoto}
\vfill

\begin{IEEEbiographynophoto}{Yejian Lyu} (yely@es.aau.dk) has been pursuing the Ph.D. in Aalborg University, Denmark since 2021. His interests in research include radio channel sounder design and channel characterization at sub-THz bands.
\end{IEEEbiographynophoto}
\vfill

\begin{IEEEbiographynophoto}{Jianhua Zhang} (jhzhang@bupt.edu.cn) is currently a professor at BUPT. Her current research interests include Beyond 5G and 6G, artificial intelligence, and data mining, especially in mmWave, THz, and massive MIMO channel modeling.
\end{IEEEbiographynophoto}
\vfill

\begin{IEEEbiographynophoto}{Gert Frølund Pedersen} (gfp@es.aau.dk) is currently a professor at Aalborg University. His research interests include radio communication for mobile terminals, especially small antennas, diversity systems, propagation, and biological effects.
\end{IEEEbiographynophoto}
\vfill

\begin{IEEEbiographynophoto}{Jonathan Borrill} (Jonathan.Borrill@anritsu.com) is currently the Chief Technology Officer (CTO) and technology leader of Anritsu.
\end{IEEEbiographynophoto}
\vfill

\begin{IEEEbiographynophoto}{Fengchun Zhang} (fz@es.aau.dk) is currently an assistant professor at Aalborg University. Her research interests are antenna array signal processing, channel characterization and over the air testing of multiple antenna systems.
\end{IEEEbiographynophoto}
\vfill

\end{document}